# A SKYRMION FLUID

G. KÄLBERMANN

Faculty of Agriculture

and

Racah Institute of Physics

Hebrew University, 91904 Jerusalem, Israel

October 28, 2018

**Abstract**

A fluid of Skyrmions coupled to the dilaton field and the $\omega$ meson field is considered. A mean field theory is developed in which the dilaton and the $\omega$ field acquire a mean value determined by the Skyrmions. The influence of the background fields on the Skyrmion profile is determined and consequently the scaling properties of the Skyrmion follow. The model obeys chiral invariance and scale invariance broken only by the dilaton trace anomaly potential. The dilaton plays the role of the scalar field as in the $\sigma + \omega$ model. The dilaton potential is augmented by terms that do not spoil the trace anomaly while allowing a fit to nuclear matter properties. The phase diagram of the model shows unique features, like a lack of solutions for certain densities and temperatures, signalling the appearance of a new phase that can not be described in terms of Skyrme baryons.

**PACS** 12.39Dc, 21.10Dr, 21.65.+f, 24.10.Pa

# 1 Introduction: Application of Skyrmions to the many body problem

Around 35 years ago Skyrme [1] proposed to describe baryons as topological solitons in a meson field theory. More recently, studies of the large $N_c$ approximation to QCD [2] suggested that indeed the meson degrees of freedom are dominant and baryons might arise as solitons [3, 8]. In the last decade there has been a large body of research on these solitons, or Skyrmions [4, 5, 6, 7]. The research has mainly focused on single baryon and few baryon systems.

Skyrme himself was interested in applying his model to systems of many nucleons aimed at the description of nuclei. After gaining support from QCD it appears that a model of nuclei based on Skyrmions would be better related to more fundamental theories of matter. This is especially appealing due to the difficulty of solving QCD explicitly.

The Skyrme model has had moderate success in dealing with the nucleon-nucleon interaction, including the attractive isoscalar central potential for which several mechanisms were proposed. It appears then natural to investigate more formal issues involving Skyrmions, such as their behavior in nuclear matter.

In the present work a mean field model of dilatons and $\omega$ mesons coupled to Skyrmions is investigated. The model is inspired in the classical $\sigma + \omega$ Walecka approach [9]. The original $\sigma + \omega$ model has been extended in order to include scale invariance by means of the dilaton field in refs.[10, 11]. In these works two scalar fields are used, the chiral partner of the pion represented by the $\sigma$ field and the dilaton. The models try to reproduce correctly both the bulk properties of nuclei and their excitation spectra. Chiral symmetry is introduced in its nonlinear realization in the former and in its linear one in the latter, while the nucleons are considered as pointlike Dirac objects. A different attempt to include chiral symmetry in the mean field theory of Walecka was done by Lynn [12]. The virtue of the Skyrme model is that it enables to include both chiral symmetry and baryons in a consistent way, the baryons being solitons of finite extent in space. These aspects make the model a good tool for the investigation of baryonic nuclear matter, but at the same time the treatment becomes more involved. We will see, however that in a dilute fluid approximation it is possible to treat the Skyrmions as essentially free particles interacting with a medium that carries the information of density and temperature. This medium acts through the mean fields of the dilaton and the $\omega$ that are determined by the Skyrmions themselves. Mishustin [13] has treated a single Skyrmion immersed in a bath of Dirac particles. Our approach is to generate the bath from the Skyrmions themselves in a self-consistent manner. Section 2 deals with the dilute fluid approximation. Section 3 describes a specific Skyrmion fluid. Section 4 treats the choice of dilaton potential. Section 5 shows the results of the mean field approximation for finite temperature and density.

## 2 Dilute Skyrmion fluid

Consider a field theory lagrangian of Skyrmions, the dilaton $\sigma$ and the $\omega$ meson [7, 14]

$$
\begin{aligned}
\mathcal{L} &= \mathcal{L}_{2\ \text{dilaton}} + \mathcal{L}_2 + \mathcal{L}_4 - V_{\text{interaction}} - V(\sigma) + \mathcal{L}_\omega \\
&= e^{2\sigma}\left[\frac{1}{2}\Gamma_0^2\, \partial_\mu\sigma\, \partial^\mu\sigma - \frac{F_\pi^2}{16}\text{tr}(L_\mu L^\mu)\right] + \frac{1}{32e^2}\text{tr}[L_\mu,\, L_\nu]^2 - g_V\ \omega_\mu B^\mu \\
&\quad - B[1 + e^{4\sigma}(4\sigma - 1)] - \frac{1}{4}(\partial_\mu\omega_\nu - \partial_\nu\omega_\mu)^2 + \frac{1}{2}e^{2\sigma}m_\omega^2\ \omega_\mu^2.
\end{aligned} \quad (1)
$$

Here

$$L_\mu \equiv U^\dagger \partial_\mu U, \quad (2)$$

where $U(\mathbf{r}, t)$ is the chiral field, $F_\pi$ is the pion decay constant and $e$ the Skyrme parameter.

The trace of the energy-momentum tensor in QCD is given by

$$T^\mu_\mu = \partial_\mu D^\mu = -\frac{9\alpha_s}{8\pi} G^a_{\mu\nu} G^{a\mu\nu} \equiv \psi^4, \tag{3}$$

where $D^\mu (= T^{\mu\nu} x_\nu)$ is the dilatation current, $\alpha_s$ is the QCD coupling constant, $G^a_{\mu\nu}$ is the gluon field, and $\psi$ is an order-parameter field—the dilaton—which represents the scalar glueball formed from the contraction of the two gluon fields; $\psi$ is taken in the fourth power in anticipation of its use as a scalar field of dimension 1. In the lagrangian of eq. (1) $\psi \approx \Gamma_0 e^\sigma$. The trace of the energy momentum tensor of eq. (1) -generated by $V_\sigma$- is constructed in accordance with eq. (3).

The lagrangian incorporates scale invariance broken only by the anomaly. All other terms are scale invariant, including the $\omega$ mass term that carries appropriate factors of $e^\sigma$.

In the present work we focus on the SU(2) version of the Skyrmion. The lagrangian possesses chiral as well as isospin symmetry. We have omitted a pion mass term that manifestly breaks chiral symmetry because it is relatively unimportant. We have also omitted $\rho$ meson coupling as it is our intention to treat symmetric nuclear matter only. The two parameters $\Gamma_0$ and $B$ are related to the glueball—or scalar field—mass through $m_\sigma^2 = 16B/\Gamma_0^2$. An accepted value of the mass is around 1.5 GeV, although the mixing of the glueball with $\pi\pi$-exchange may bring it down to the below GeV range.(A critique to the use of the dilaton in nuclear matter applications can be found in ref. [15].)

Analogously to the mean field theory of pointlike baryons, we consider an ensemble of essentially free Skyrmions. Each Skyrmion will be accompanied by its own dilaton and $\omega$ fields. Although we deal with free Skyrmions, the average properties of the ensemble are still included in distribution functions that depend on the density and temperature.

For the $N$ Skyrmions we use the product ansatz

$$U_{B=N}(\mathbf{r}, \mathbf{R}_1, \mathbf{R}_2, \cdots, \mathbf{R}_N) = U(\mathbf{r} - \mathbf{R}_1) U(\mathbf{r} - \mathbf{R}_2) \cdots U(\mathbf{r} - \mathbf{R}_N), \qquad (4)$$

whereas for the scalar fields we use an additive ansatz

$$\begin{aligned} \sigma_{B=N} &= \sigma_1 + \sigma_2 + \cdots + \sigma_N, \\ &= \sigma_0 + \delta\sigma_1 + \delta\sigma_2 + \cdots + \delta\sigma_N, \qquad (5) \\ \omega_{B=N} &= \omega_1 + \omega_2 + \cdots + \omega_N, \\ &= \omega_0 + \delta\omega_1 + \delta\omega_2 + \cdots + \delta\omega_N, \qquad (6) \end{aligned}$$

Where $\sigma_0, \omega_0$ are the mean field constant values of the dilaton and the $\omega$ and $\delta\sigma, \delta\omega$ represent the fluctuations. The fields $\sigma, \delta\sigma, \omega, \delta\omega$ depend on the same arguments as the Skyrmion they are attached to. The ansätze above insure total baryon number $= N$ and allow an easy separation of the mean field degrees of freedom from the microscopic excitations inside each Skyrmion. In thermal equilibrium, the mean field fields will depend on the temperature $T$ and the chemical potential $\mu$. Self-consistency then requires that the values of $\sigma_0$ and $\omega_0$ have to be determined by the properties of the ensemble. For a certain distribution function $\mathbf{f}(\mu, T)$ the temperature, we then demand the phase space thermal average to be

$$\begin{aligned} <\sigma> &= <\sigma_0 + \delta\sigma_1 + \cdots + \delta\sigma_N> \\ &= (2\pi)^{-3N} \int d\vec{R}_1 d\vec{P}_1 \cdots d\vec{R}_N d\vec{P}_N \; \mathbf{f} \;\; \sigma \\ &= \sigma_0 \qquad (7) \end{aligned}$$

with a similar equation for $\omega$. The mean field average value of the fluctuations vanishes by definition. In eq. (7) the integration is over the collective coordinate coordinates $\vec{R}_i$ of the Skyrmions and their conjugated momenta.

In the diluted fluid approximation, Skyrmion interactions are neglected and consequently there are no potential terms depending on the coordinates $\mathbf{R}_i$ of each baryon. There remain only kinetic energy terms for each Skyrmion individually. Standard quantization of these terms will then determine the corresponding wave functions to be plane waves, as for the conventional Fermi gas model.

A crucial point in the mean field theory of skyrmions is the topological baryon density

$$B^\mu = \frac{\epsilon^{\mu\alpha\beta\gamma}}{24\pi^2} \mathrm{tr}\left[\left(U^\dagger \partial_\alpha U\right)\left(U^\dagger \partial_\beta U\right)\left(U^\dagger \partial_\gamma U\right)\right], \tag{8}$$

where $\epsilon^{\mu\alpha\beta\gamma}$ is the totally antisymmetric tensor density, and it is easily seen that

$$\partial_\mu B^\mu = 0. \tag{9}$$

The corresponding baryon number becomes

$$B_0 = \frac{\epsilon^{ijk}}{24\pi^2} \mathrm{tr}\left[\left(U^\dagger \partial_i U\right)\left(U^\dagger \partial_j U\right)\left(U^\dagger \partial_k U\right)\right], \tag{10}$$

Using the product ansatz of eq. (4) in the baryon number above allows us to calculate the mean field average by integrating over the Skyrmions' coordinates and momenta. Consistently with the approximation of a fluid of free Skyrmions we demand that there is no overlap between the Skyrmions profile functions, and further assume that the profile drops to zero at a distance smaller than the inter-Skyrmion separation. This condition is needed in order to insure integer baryon number within each Skyrmion cell, because the single baryon density depends on the values of the profile function at the particles' location and at infinity. In the present case, infinity is replaced by a finite distance. Typically, this distance is of the order of $0.8\ fm$, whereas the interparticle separation in normal nuclear matter is approximately $2\ fm$. It is then not very unrealistic to assume that most of the topological baryon density is contained within a volume smaller than the average volume available to each Skyrmion. Note that, even if there is some overlap between the Skyrmions, the choice of product ansatz insures total baryon number $= N$. To this approximation we obtain

$$\begin{aligned} B_0 &= b_1 + b_2 + \cdots + b_N, \\ <B_0> &= N/V \end{aligned} \qquad (11)$$

where $b_i$ is the baryon density of the $i^{th}$ Skyrmion, V is the volume of the fluid, and the average is defined in eq. (7). The integration over the collective coordinates of the Skyrmions is mathematically equivalent to an integration over the coordinates **r** in eq. (4), this is the reason why the averaging above essentially counts the number of Skyrmions. After the Fermi averaging, the localized baryon density becomes uniformly smeared out. There is no residue of the individual Skyrmion densities. We will see below that the influence of the mean fields on the Skyrmions operates solely through the dilaton field. Although the baryon density of each Skyrmion will vary accordingly, the integrated density is scale invariant and will still contribute the value $B = 1$ for each Skyrmion. The dilute fluid approximation is therefore closely related to the mean field Dirac model with two major differences: 1) there is a different dynamics for the baryons dictated by the lagrangian of eq. (1) and, 2) there is a clear way to uncover the baryon response to the medium.

## 3    Mean field Skyrmion fluid

The many body Skyrmion fluid in the dilute approximation is built from the product ansatz of eq. (4) for the Skyrmions and the additive ansatz for the mesons of eq. (5). We therefore first focus on the single baryon case. Each soliton is constructed as a static solution of the equations of motion. The collective coordinate energy of the Skyrmion translation is later introduced in an adiabatic approximation. The single baryon ansätze are then

$$\begin{aligned} U(\mathbf{r}) &= \exp[i\boldsymbol{\tau}\cdot\hat{\mathbf{r}}F(r)], \\ \omega^\mu(\mathbf{r}) &= (\omega(r),0,0,0) \end{aligned} \qquad (12)$$

Substituting eq. (12) into eq. (1) we find the static mass of the Skyrmion

$$\begin{aligned} M &= 4\pi\int_0^\infty r^2\,dr\,M(r) \\ M(r) &= e^{2\sigma}\frac{F_\pi^2}{8}\left[F'^2+2\frac{\sin^2 F}{r^2}\right]+\frac{1}{2e^2}\frac{\sin^2 F}{r^2}\left[\frac{\sin^2 F}{r^2}+2F'^2\right]+V(\sigma) \\ &\quad +e^{2\sigma}\frac{1}{2}\Gamma_0^2\,\sigma'^2-\frac{1}{2}\omega'^2+\frac{g_V}{2\pi^2 r^2}\frac{\omega\;F'\;\sin^2 F}{}-\frac{1}{2}m_\omega^2\,\omega^2\,e^{2\sigma} \qquad (13) \end{aligned}$$

where primes denote derivatives with respect to $r$. The Euler-Lagrange equations for the static profile and the mesons become

$$\begin{aligned} \left(e^{2\sigma}+\frac{8\sin^2 F}{\tilde{r}^2}\right)F''&+2e^{2\sigma}F'(1/r+\sigma')+\frac{4\sin 2F F'^2}{\tilde{r}^2}-\frac{e^{2\sigma}\sin 2F}{r^2} \\ &\quad -\frac{4\sin^2 F\sin 2F}{r^2\tilde{r}^2}+\frac{2g_V\,\omega'\,\sin^2 F}{\tilde{r}^2}=0 \\ \Gamma_0^2\;e^{2\sigma}\left(\sigma''+2\sigma'^2\right.&\left.+\frac{2\sigma'}{r}\right)-\frac{F_\pi^2 e^{2\sigma}}{4}\left(F'^2+\frac{2\sin^2 F}{r^2}\right) \\ &\quad -\frac{dV_\sigma}{d\sigma}+m_\omega^2\,\omega^2\,e^{2\sigma}=0 \\ \omega''+\frac{2\omega'}{r}&-m_\omega^2\,\omega\,e^{2\sigma}+\frac{g_V F'\sin^2 F}{2\pi^2 r^2}=0 \qquad (14) \end{aligned}$$

where $\tilde{r}=eF_\pi r$ and again primes denote derivatives with respect to $r$.

Using eq. (5) and the fact that the fluctuation of $\sigma$ and $\omega$ vanishes in the mean field state we readily find that except for small contributions coming from the $\sigma$ potential the equations above are modified in the presence of a nonvanishing $\omega_0$ and $\sigma_0$ by the simple scaling law

$$r\to e^{-\sigma_0}r\,,\qquad \omega\to e^{\sigma_0}\omega \qquad (15)$$

It is then unnecessary to solve the equations of motion for the single Skyrmion with the meson fluctuations included. It suffices to solve them for a free Skyrmion and then rescale the $\omega$ field and the radial distance. Moreover, for dilaton masses of the order of 1 GeV and more, the fluctuation $\delta\sigma$ is altogether negligible. Therefore the scaling law becomes exact. Analogously it is easy to show that the static mass of eq. (13) scales as

$$M = M_0 e^{\sigma_0} \tag{16}$$

where $M_0$ is the mass for $\sigma = 0$.

Before proceeding to the many body case, we consider the Lorentz boosting of the static Skyrmions. We perform a boost on the collective coordinate $R(t)$ of each Skyrmion. For the sake of simplicity consider a Lorentz boost along the $x$ axis with velocity parameter $v$:

$$
\begin{aligned}
x \to \tilde{x} &= \frac{x - R(t)}{\sqrt{1-v^2}} \\
\tilde{y} &= y \\
\tilde{z} &= z \\
F(\vec{\mathbf{r}}) &\to F(\vec{\tilde{\mathbf{r}}}) \\
\sigma(\vec{\mathbf{r}}) &\to \sigma(\vec{\tilde{\mathbf{r}}}) \\
\omega^\mu &\to \left( \frac{\omega(\vec{\tilde{\mathbf{r}}})}{\sqrt{1-v^2}}, \frac{v\omega(\vec{\tilde{\mathbf{r}}})}{\sqrt{1-v^2}}, 0, 0 \right) \\
B^\mu &\to \left( \frac{B_0(\vec{\tilde{\mathbf{r}}})}{\sqrt{1-v^2}}, \frac{vB_0(\vec{\tilde{\mathbf{r}}})}{\sqrt{1-v^2}}, 0, 0 \right)
\end{aligned}
\tag{17}
$$

with $\omega^\mu B_\mu = \omega_0(\vec{\tilde{\mathbf{r}}}) B_0(\vec{\tilde{\mathbf{r}}})$. Introducing the above transformation in eq. (1) and calculating the Hamiltonian, we find the energy of a Skyrmion in motion to be

$$E_p = \left( E_2 + E_\sigma - E_\omega \right) \frac{2p^2 + 3M^2}{3\epsilon\, M} + E_4 \frac{4\, p^2 + 3M^2}{3\epsilon\, M} + \frac{M}{\epsilon} \left( U_\sigma - U_\omega + U_{int} \right) \tag{18}$$

where
$$\epsilon = \sqrt{p^2 + M^2}, \quad p = \frac{Mv}{\sqrt{1-v^2}},$$
$M$ is the static mass of eq. (13) for nonvanishing $\omega_0, \sigma_0$ and

$$\begin{aligned}
E_2 &= \frac{4\pi F_\pi^2}{8} \int r^2 dr\; e^{2\sigma} \left(F'^2 + \frac{2\sin^2 F}{r^2}\right) \\
E_4 &= \frac{4\pi}{2e^2} \int dr \left(2F'^2 + \frac{\sin^2 F}{r^2}\right) \\
E_\sigma &= \frac{4\pi \Gamma_0^2}{2} \int r^2 dr\; e^{2\sigma} \sigma'^2 \\
E_\omega &= 4\pi \int r^2 dr\; \omega'^2 \\
U_\sigma &= 4\pi \int r^2 dr\; V_\sigma \\
U_\omega &= 4\pi \int r^2 dr\; e^{2\sigma} \frac{m_\omega^2 \omega^2}{2} \\
U_{int} &= \frac{2g_V}{\pi} \int dr\; \omega\, F'\, \sin^2 F
\end{aligned} \quad (19)$$

The energy $E_p$ of eq. (18) enters the single particle distribution functions

$$\begin{aligned}
n_p &= \left(\exp\left[\frac{\epsilon_p - \mu}{kT}\right] + 1\right)^{-1} \\
\bar{n}_p &= \left(\exp\left[\frac{\bar\epsilon_p + \mu}{kT}\right] + 1\right)^{-1}
\end{aligned} \quad (20)$$

where $\epsilon_p = E_p + g_V\,\omega_0, \quad \bar\epsilon_p = E_p - g_V\omega_0$.

We can now write down the energy of N Skyrmions per unit volume in the mean field approximation for symmetric nuclear matter.

$$E_V = 4\int \frac{d^3 p}{(2\pi)^3}\; E_p(n_p + \bar n_p) + V_\sigma(\sigma_0) - \frac{1}{2} e^{2\sigma_0} m_\omega^2 \omega_0^2 + g_V\,\omega_0\,\rho_V \quad (21)$$

where
$$\rho_V = 4\int \frac{d^3 p}{(2\pi)^3}\; (n_p - \bar n_p) \quad (22)$$

At $T = 0$ we have $n_p = \Theta(p_F - p), \bar n_p = 0$ yielding the equations of motion for the mean fields

$$0 = \frac{\partial E_V}{\partial \sigma_0} = 4\int \frac{d^3p}{(2\pi)^3} \frac{\partial E_p}{\partial \sigma_0} + \frac{dV_\sigma}{d\sigma_0} - m_\omega^2 e^{2\sigma_0}\omega_0^2$$

$$0 = \frac{\partial E_V}{\partial \omega_0} = m_\omega^2 e^{2\sigma_0}\omega_0 - \frac{2g_V p_F^3}{3\pi^2} \quad (23)$$

At finite temperature we have to maximize the pressure instead of minimizing the energy. The contribution to the pressure of a single Skyrmion is given by

$$P_p = \left(E_2 + E_\sigma - E_\omega\right)\frac{2p^2 - 3M^2}{9\epsilon\, M} + E_4\frac{4p^2 + 3M^2}{9\epsilon\, M} - \frac{M}{\epsilon}\left(U_\sigma - U_\omega\right) \quad (24)$$

and the pressure of the ensemble per unit volume becomes

$$P_V = 4\int \frac{d^3p}{(2\pi)^3}\, P_p(n_p + \bar{n}_p) - V_\sigma(\sigma_0) + \frac{1}{2}e^{2\sigma_0}m_\omega^2\omega_0^2 \quad (25)$$

As mentioned above, due to the large dilaton mass one can neglect the fluctuation of the dilaton field inside the Skyrmion. Numerical evaluation of $\delta\sigma$ for a dilaton mass above 1 GeV indeed gives a negligible dilaton field $\delta\sigma \approx 0$. Moreover, the simplest approximation to the mean field demand of a vanishing expectation value of the fluctuation of the $\omega$ field is $\delta\omega = 0$. This first approximation to the problem permits a large simplification of the single particle's energy and pressure. Using the virial theorem for the soliton profile appropriate for the case of vanishing dilaton and $\omega$ fluctuation, $E_2 = E_4 = \frac{1}{2}M$, we obtain

$$M = \pi F_\pi^2 e^{2\sigma_0}\int r^2 dr\left(F'^2 + \frac{2\sin^2 F}{r^2}\right) = e^{\sigma_0}M_0$$

$$\epsilon = E_p = \sqrt{p^2 + e^{2\sigma_0}M_0^2}$$

$$P_p = \frac{p^2}{3E_p} \quad (26)$$

To this approximation the mean field Skyrmion fluid becomes identical to the Dirac mean field approach, with the bonus of knowing how to calculate the reaction of the single Skyrmion to the bath using the scaling properties determined by the dilaton. As it is our goal to investigate the most important contribution to the Skyrmion properties due to the bath, we will proceed along this line and neglect the dilaton and the $\omega$ meson fluctuations inside the Skyrmion. The results we will obtain will then be applicable to a conventional mean field theory of Dirac pointlike particles coupled to the dilaton and the $\omega$ meson. The key new ingredient in such a treatment would then be the use of a single scalar field in order to fit the nuclear matter properties.

## 4 The dilaton potential

In eq. (1) we introduced the conventional glue potential that reflects the trace anomaly [16], where the 'bag constant' is $B \approx (240 \text{ MeV})^4$. The potential can be supplemented by other terms that do not affect the anomaly. Modifications of the potential have been proposed in order to include chiral symmetry breaking corrections and quark contributions to the trace anomaly [11], whereas Furnstahl et al. [10] opted for an approximation to the potential that takes into account anomalous dimensions acquired by the scalar fields upon renormalization. In any event there is a need to modify the potential in order to fit the properties of nuclear matter. The difficulty arises mostly in fitting the low value of the nuclear compressibility modulus $\kappa \approx 270$ MeV. In the abovementioned works two scalar fields come into play, the dilaton and a scalar partner of the pion, the $\sigma$ meson. In the present approach there is no room for the latter, because we are working in the nonlinear realization of chiral symmetry due to the use of the Skyrme lagrangian. We therefore allow for modifications of the dilaton potential that do not affect the anomaly, but influence the predictions for nuclear matter. Terms of the form $e^{n\sigma} - 1$ , generically referred to as no-log terms [11], can be added to the potential leaving the vacuum expectation value of the $\sigma$ to be that dictated by the anomaly potential, chosen here to be $\sigma = 0$ by a simple shift of the original dilaton field. The actual choice of the functional form of the new terms in the potential turns out to be relatively unimportant. (There exists also the possibility of modifying the $\omega$ potential as in ref. [11], introducing higher order terms in the $\omega$ field).

Let us consider the constraints on $V_\sigma$ demanded by nuclear matter phenomenology. The potential has to be such that at the saturation density of nuclear matter $\rho_0 = .154$ baryons/fm$^3$: a) the binding energy per nucleon is 16 MeV ; b) the binding energy is maximal ; c) the compressibility is of the order of 270 MeV; d) the dilaton and $\omega$ fields obey the mean field equations. These conditions translate into the following equations

$$\omega_0 = \frac{g_V \, \rho_0 \, e^{-2\sigma_0}}{m_\omega^2}$$

$$\frac{\partial V_\sigma}{\partial \sigma_0} - \omega_0^2 m_\omega^2 e^{2\sigma_0} + Q = 0 \qquad (27)$$

where

$$Q = \frac{M^2 p_F^2}{\pi^2}\left[\sqrt{1+z^2} - \frac{1}{2}z^2 \ln\left(\frac{\sqrt{1+z^2}+1}{\sqrt{1+z^2}-1}\right)\right],$$

$$p_F = \left(\frac{3}{2}\pi^2 \rho\right)^{1/3}, z = \frac{M}{p_F}.$$

$$E_V = V_\sigma + \frac{m_\omega^2 \omega_0^2}{2} + \frac{1}{4}Q + \frac{p_F^3 \sqrt{p_F^2 + M^2}}{2\pi^2}$$
$$= \rho_0(M_0 - 16 \text{ MeV})$$
$$\frac{E_V}{\rho_0} = \sqrt{p_F^2 + M^2} + \frac{g_V^2 \rho_0 e^{2\sigma_0}}{m_\omega^2}. \qquad (28)$$

$$\kappa = 9\rho_0 \left[\frac{\partial^2 E_V}{\partial \rho^2} - \left(\frac{\partial^2 E_V}{\partial \rho \partial \sigma}\right)^2 \left(\frac{\partial^2 E_V}{\partial \sigma^2}\right)^{-1}\right] \qquad (29)$$

where

$$\begin{aligned}
\frac{\partial^2 E_V}{\partial \rho^2} &= \frac{p_F^2}{3\rho_0\sqrt{p_F^2+M^2}} + \frac{g_V^2}{m_\omega^2 e^{2\sigma_0}} \\
\frac{\partial^2 E_V}{\partial \rho \partial \sigma} &= \frac{M^2}{\sqrt{p_F^2+M^2}} - \frac{2g_V^2 \rho_0}{m_\omega^2 e^{2\sigma_0}} \\
\frac{\partial^2 E_V}{\partial \sigma^2} &= \frac{d^2 V_\sigma}{d\sigma^2} + 4Q - \frac{2p_F^3 M^2}{\pi^2 \sqrt{p_F^2+M^2}} - \frac{2g_V^2 \rho_0^2}{m_\omega^2 e^{2\sigma_0}}
\end{aligned} \quad (30)$$

In order to fulfill the above constraints we introduced an additional piece in the dilaton potential with three new free parameters that proved to be quite successful in reproducing the nuclear matter phenomenology

$$V_\sigma = B[1+e^{4\sigma}(4\sigma-1)] + B\left[a_1(e^{-\sigma}-1) + a_2(e^\sigma-1) + a_3(e^{2\sigma}-1) + a_4(e^{3\sigma}-1)\right] \quad (31)$$

where B is fixed and the anomaly condition requires

$$\frac{dV_\sigma}{d\sigma} = 0 \quad (32)$$

at $\sigma = 0$ implying $a_1 = a_2 + 2a_3 + 3a_4$.

We have chosen the term multiplied by $a_1$ to have a negative power of $\sigma$ in order to avoid the introduction of a second minimum in the potential for $\sigma < 0$. The only sensible minimum then remains the one at $\sigma = 0$. There is a need to introduce three new parameters in order to reproduce correctly the nuclear matter binding energy at the right density, and the compressibility factor.

The potentials for the fluctuations of the $\sigma$ and $\omega$ fields are then determined by the averages of eq. (7) to be to lowest order a massterm interaction

$$V(\delta\sigma) = B\frac{1}{2}\delta\sigma^2 \left[16e^{4\sigma_0}(1+4\sigma_0) + a_1 e^{-\sigma_0} + a_2 e^{\sigma_0} + 2a_3 e^{2\sigma_0} + a_4 \frac{9}{2}e^{3\sigma_0}\right]$$

$$V(\delta\omega) = \frac{1}{2}e^{2\sigma_0}m_\omega^2\delta\omega^2 \qquad (33)$$

The effective mass of the dilaton is not fixed, because it depends on the parameter $\Gamma_0$ of eq. (1), that has to be prescribed separately. The mass will enter only in finite nuclei calculations.

## 5 Results and discussion

The conditions on the dilaton potential of eqs. (27, 28 , 29) were implemented by choosing a suitable effective mass at the nuclear matter saturation density and a certain compressibility. We took $M^* = 694$ MeV corresponding to $\sigma_0 = -0.3$ and $\kappa = 300$ MeV. The value of $\kappa$ we used is very close to the measured one [18] - although much lower values are suggested by the analysis of Blaizot [19]- ,whereas for the effective mass we took a conservative figure, although many authors tend to select a lower value of around 600 MeV. We have found that there are no qualitative differences in the predictions of the model when the latter mass is used. In all the fits we kept the bag constant at $B = (240 \text{ MeV})^4$, although we were able to fit the data wit a large range of values of this parameter.

With the above choice we find : $a_1 = $ -5.53, $a_2 = $ -54.74, $a_3 = $ 48.63, $a_4 = $ 16.01 and, $g_V = $ 7.29. Although it may seem that the dilaton potential coefficients are a bit large, one has to bear in mind that they multiply expressions very close to zero for all the attainable values of $\sigma$. The $\omega$ meson coupling constant turns out to be smaller than expected but still within the range quoted in the literature. Other terms in the $\omega$ meson potential might be needed in order to allow higher values of $g_V$.

Figure 1 shows the binding energy per nucleon as a function of the density both for the *normal* and *abnormal* solutions. The normal branch possesses all the desired requirements for nuclear matter, therefore at $T = 0$ it represents the physically stable phase. The existence of an abnormal branch is well known. It arises from the functional dependence of the scalar field potential. This branch is characterized by large negative values of the scalar field. In some models the normal and abnormal branches cross each other at some critical density, suggesting a transition to a dense, chiral symmetry restored phase, accompanied by a dramatic decrease in the nucleon effective mass. Chiral symmetry is built-in in the Skyrmion fluid, hence, the effective nucleon mass is not an indicator of chiral symmetry restoration. The abnormal branch is here totally spurious, and might be possibly reached in a metastable situation.

At finite temperature the criterion of maximal pressure for a fixed chemical potential determines the physical phase. Again, the *normal* branch prevails.

Figure 2 shows the nucleon effective mass as a function of density and temperature. (The apparent cusp in figure 2 is due to the rudimentary plotting procedure).

In the Walecka model -and many others similar to it- the nucleon mass decreases as a function of density [17]. This is not the case in the Skyrmion fluid model. The minimum effective mass arises here at a density of around 1.7 $\rho_0$. Contrarily, the usual parametrizations of the scalar field potential in Walecka models, do not show this limitation. The reason for this difference can be traced back to the dynamics dictated by the dilaton, especially to the modified trace anomaly potential.

The scalar field potential of the Walecka type models is built in terms of powers of the scalar field, up to fourth order generally, to satisfy renormalizability requirements. (Although this condition is usually hard to meet due to the need for a negative coefficient for the fourth order term in most successful fits.) However, the dilaton potential is made to obey the condition demanded by the trace anomaly. Only no-log terms that do not spoil the anomaly are allowed. This strong demand makes the potential qualitatively different. The solutions to the nuclear matter equations with this potential for the normal branch, exist only in a narrow range of values of $\sigma$ near $\sigma = 0$.

It is now possible to understand why the effective mass increases beyond a certain density: The dilaton attractive contribution to the mass is limited, while the $\omega$ meson repulsion grows in direct proportion to the nuclear matter density, eq. (27). The solution of the nuclear matter equations then tend to push the dilaton towards positive values in order to fulfill the scaling properties of the model, eq. (15).

The Skyrmion properties are essentially determined by the scaling characteristics of the model. The mass of figure 2 scales as $e^{\sigma_0}$ whereas the root mean square radius scales as $e^{-\sigma_0}$. The Skyrmion swells as a function of density up to densities below $1.7\rho_0$. This is due to the attractive effect induced by the dilaton. Above a certain density the Skyrmion starts to shrink, due to the abovementioned repulsive effect in a manner independent of temperature. This is reminiscent to the behavior of the Skyrmion interaction at short range (high density) that is known to be repulsive and strong. Other properties, like $g_A$, magnetic moments, etc, follow directly from the scaling of eq. (15).

Figures 3-5 depict the pressure, chemical potential and entropy as a function of density and temperature. Although it may appear that there could be a phase transition between $T = 125$ MeV and $T = 175$ MeV, calculations performed with a finer grid than the one shown in the graphs, does not support such a scenario. The transition between those temperatures is smooth.

At a temperature of 25 MeV the graphs of the pressure and chemical potential show a rough behavior due to the proximity of this temperature to the liquid-gas phase transition that appears in a broad class of Walecka type models [9].

The entropies of figure 5 were calculated with three different methods. The agreement between the numbers obtained determine the accuracy of the minimization procedure to be better than 0.5 %.

As shown in figure 6, above $T = 190$ MeV there cease to exist solutions around the the normal density of nuclear matter. Above $T = 220$ MeV solutions cease to exist for densities below $4\rho_0$, perhaps signalling the appearance of a different phase that we can not describe in terms of Skyrmions and mesons. Other degrees of freedom must come into play, perhaps quarks and gluons. This behavior is different as compared to the results of conventional models [17] for which there are solutions at all densities and temperatures with a phase transition to a baryonic plasma phase above a certain temperature typically of the order of 200 MeV. The reason for this difference can be traced back to the special features of the dilaton coupling and potential.

In the present work we have paved the way for a treatment of a Skyrmion fluid based on the principles of scale invariance and chiral invariance. In a later work we will test the approach and especially the potential obtained in finite nuclei calculations in the Thomas-Fermi approach as well the corrections due to the rotational degrees of freedom of the Skyrmion. In a finite nucleus calculation new parameters will enter, like the dilaton mass. It is then a bit risky to venture any predictions for such a system. A much more complicated and ambitious task would be to tackle the introduction of Skyrmion interactions.

Acknowledgements

It is a pleasure to thank Prof. J.M. Eisenberg for his constructive criticism, Mr. Reem Sari for his help in drawing the graphs and Ms. Shula Coster for her Tex guidance. This work was supported in part by the Israel Science Foundation.

**Figure Captions**

Fig. 1: Binding energy per nucleon in MeV as a function of $\rho/\rho_0$. Normal branch, dashed line and abnormal branch, full line.

Fig. 2: Effective Skyrmion mass $M^*$ in MeV as a function of $\rho/\rho_0$. Full line, $T = 25$ MeV, dashed line $T = 75$ MeV, dash-dot line, $T = 125$ MeV and dotted line, $T = 175$ MeV.

Fig. 3: Pressure in MeV/fm$^3$ as a function of $\rho/\rho_0$. Line definitions as in Fig. (2).

Fig. 4: Chemical potential in MeV as a function of $\rho/\rho_0$. Line definitions as in Fig. (2)

Fig. 5: Entropy per baryon as a function of $\rho/\rho_0$. Line definitions as in Fig. (2)

Fig. 6: Pressure in MeV/fm$^3$ as a function of $\rho/\rho_0$. Dotted line, $T = 190$ MeV, full line, $T = 200$ MeV and dashed line, $T = 220$ MeV.

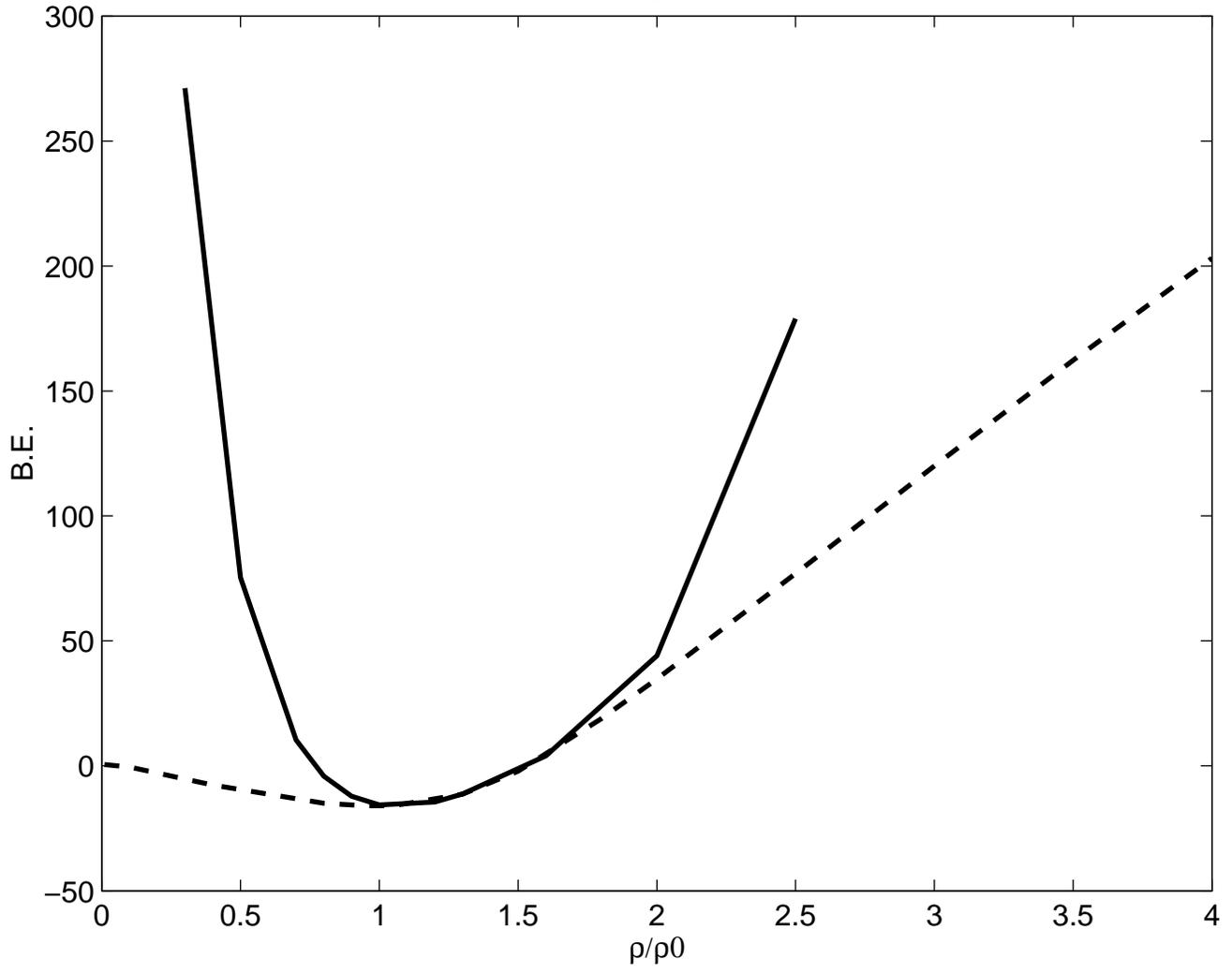

Fig. 1

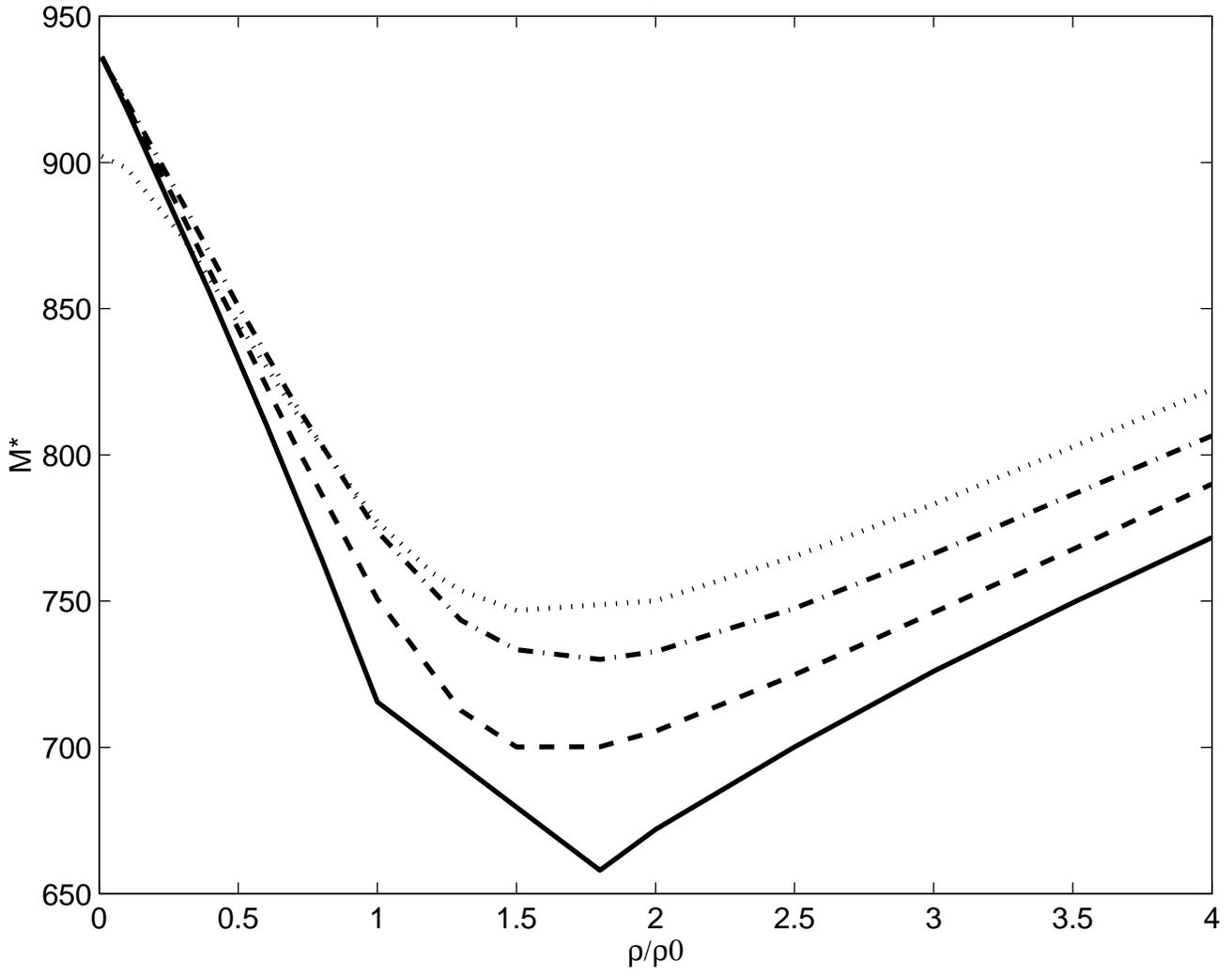

Fig. 2

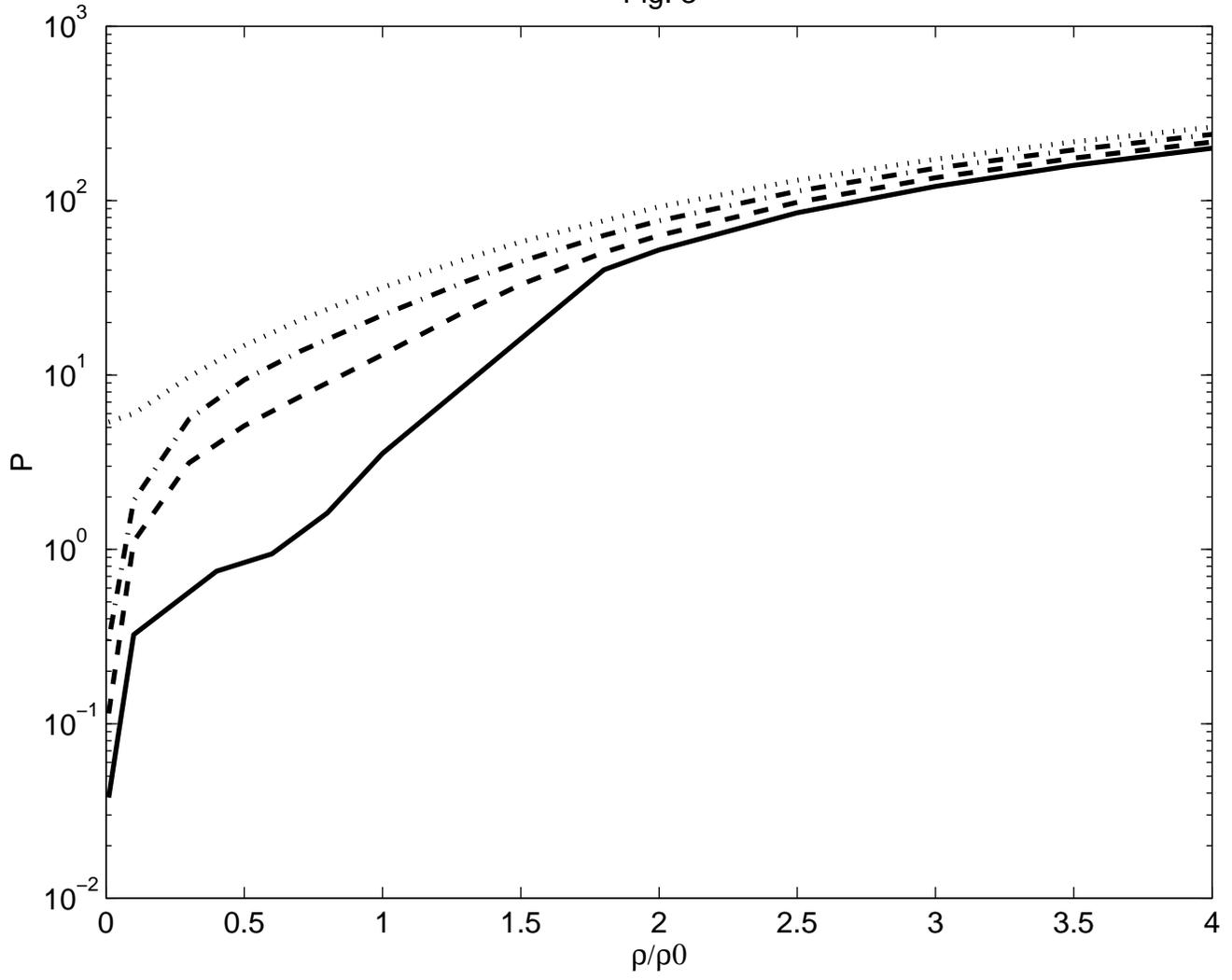

Fig. 3

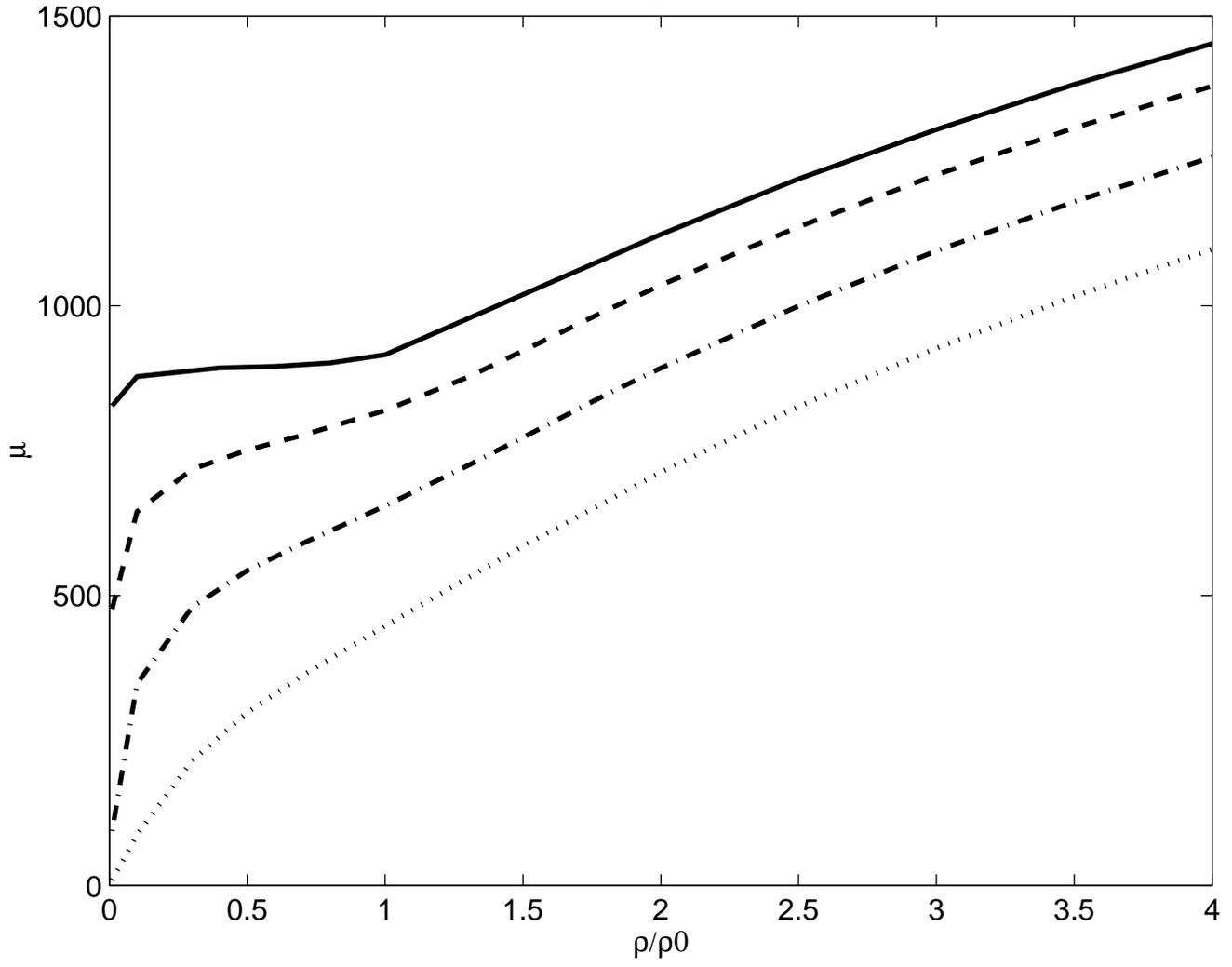

Fig. 4

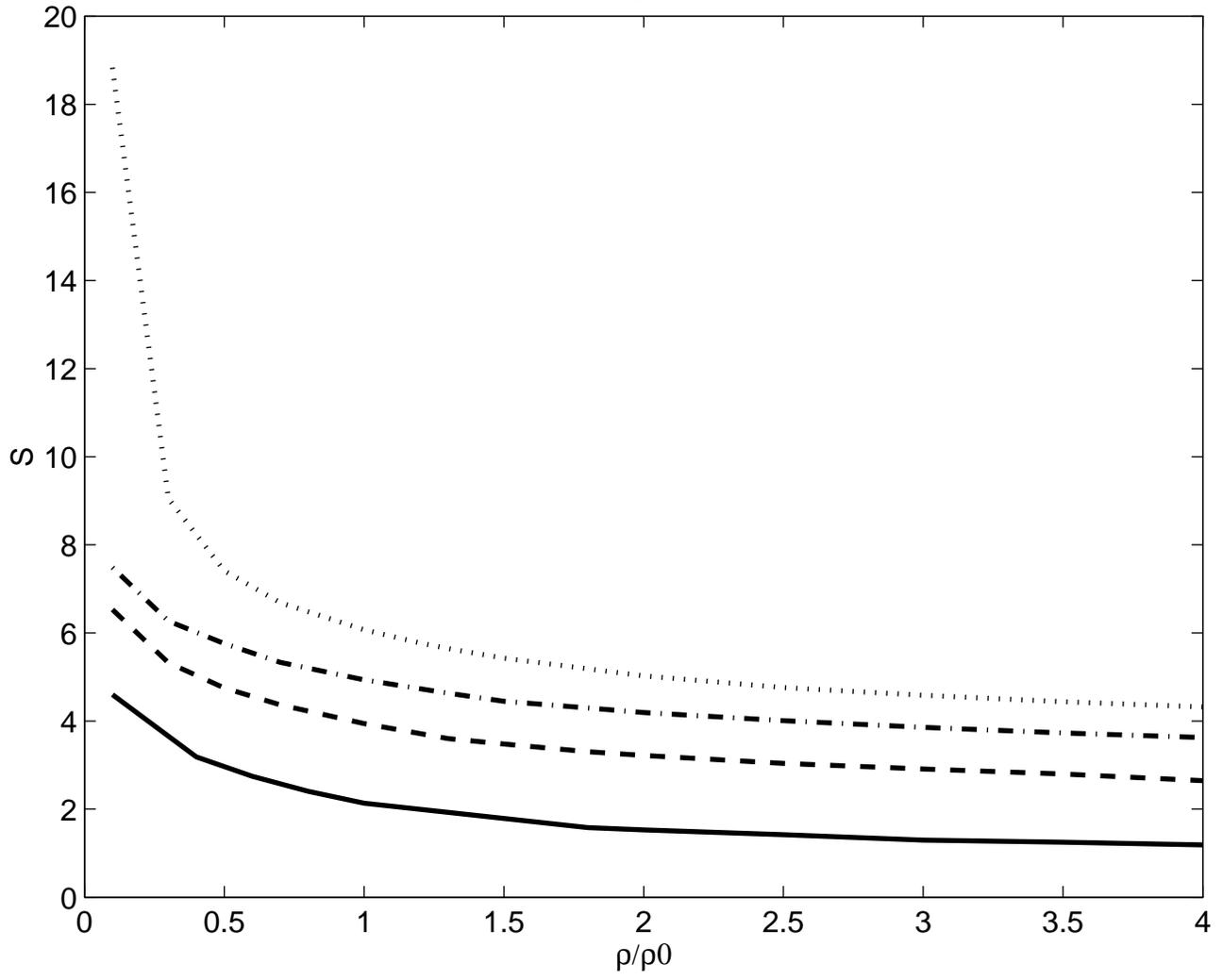

Fig. 5

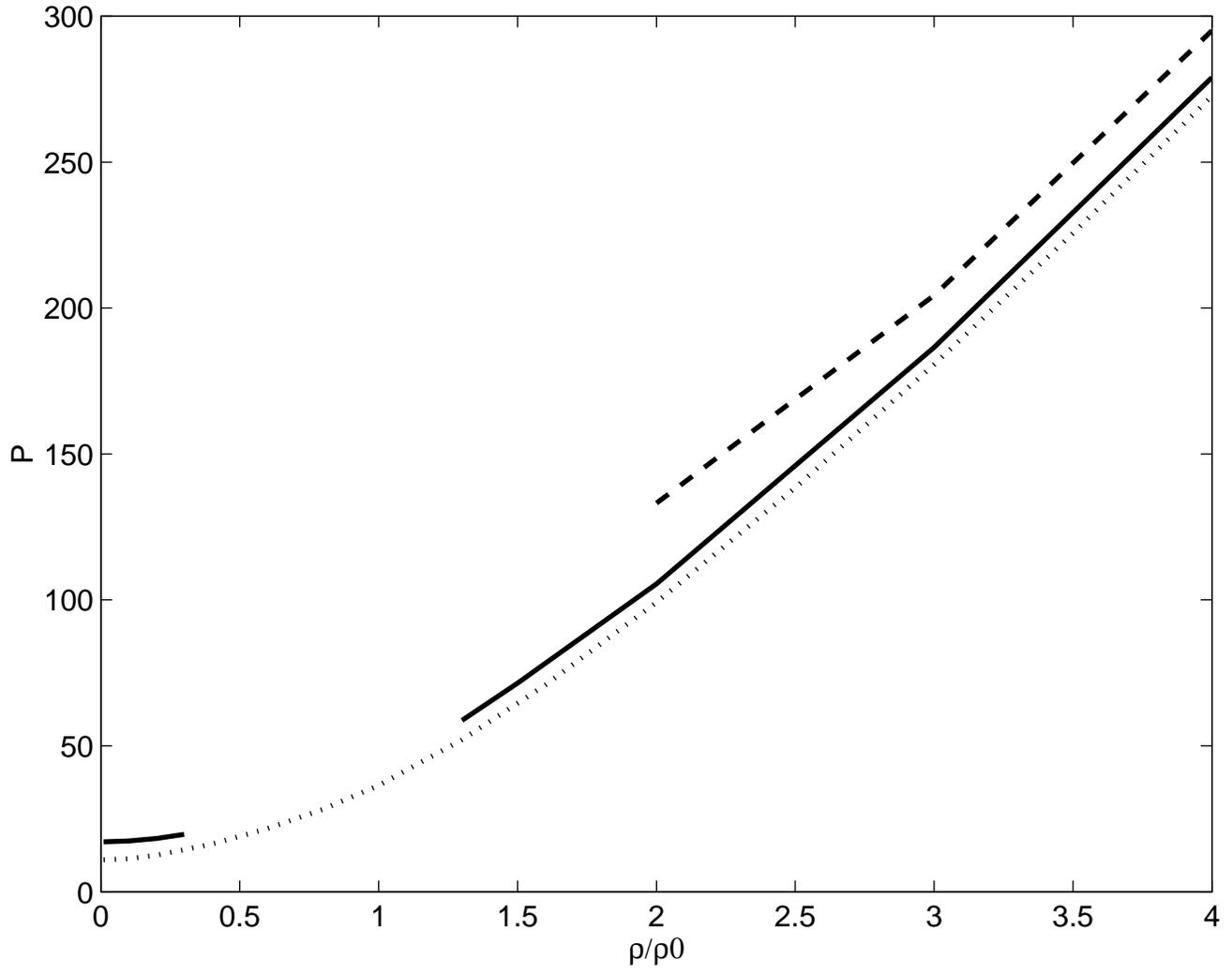

Fig. 6